**Title:**

# From Closed-Loop Optimization to Open Decision Making: Coupled Digital Twins for Predictive and Autonomous Microscopy

**Authors:**

Yu Liu[1*], Boris Slautin[1], Ian Mercer[2], Jon-Paul Maria[2], and Sergei V. Kalinin[1*]

[1] Department of Materials Science and Engineering, University of Tennessee, Knoxville, Tennessee, 37996, USA

[2] Materials Science and Engineering Department, Materials Research Institute, the Pennsylvania State University, University Park, Pennsylvania, 16802, USA

* Corresponding author: yliu206@utk.edu, sergei2@utk.edu

## Abstract

Automated experimentation is moving from closed-loop optimization toward open decision-making, where human or AI planners must forecast the consequences of candidate actions before executing them. Such forecasts require a model of both sides of the experiment: how the sample is likely to respond and what the instrument is likely to detect. We therefore introduce a coupled digital-twin framework that separates these roles and then links them. In this framework, the sample twin encodes material state inferred from prior knowledge and measurements till the moment. The instrument twin captures signal formation, feedback dynamics, and operating constraints based on prior knowledge. When coupled, the two twins estimate expected outcomes, uncertainty, and risk for candidate microscope operations. For amplitude-modulation scanning probe microscopy, we realize this framework with a physics-informed encoder of force-distance curves, a deterministic scanner model of cantilever and feedback dynamics, and sparse learned residual corrections. The encoder first recovers scanner-driving descriptors with sub-nanometer accuracy. The calibrated scanner then reproduces typical traces within a few nanometers and identifies operating-point noise amplification as the main source of mismatch. Supplementary phase analysis localizes residual error to the phase channel, which clarifies where added physics is needed. Together, these results establish coupled sample and instrument twins as a practical foundation for predictive microscope operation and autonomous experimental planning.

## I. Introduction

Artificial intelligence, machine learning, and theory-guided materials prediction are changing how physical experiments are designed. This shift is visible at several scales, from individual instrument automation[1-7] to distributed self-driving laboratories[8-11] and workflows that connect discovery to manufacturing[12-14]. Despite this breadth, much of the field still follows a closed-loop optimization model. In that model, the instrument updates control parameters to maximize a predefined reward by constructing surrogate model of prediction and uncertainty within a selected parameter space. Bayesian optimization and related strategies have made this model efficient in microscopy, synthesis, reaction optimization, and materials discovery.

A natural next step is the transition from closed-loop optimization toward open decision-making. In this regime, a human operator or AI planner must choose among measurement modalities, operating conditions, locations, and action sequences under evolving uncertainty[15-18]. The task is therefore broader than finding the best setting for a fixed objective. Before an experiment is run, a useful planner must predict likely measurement outcomes, estimate uncertainty, evaluate risk, and identify which action will be most informative. This predicted outcomes in different forms (next step vs. finite horizon vs. full roll-outs) are a foundation for optimization approaches such as model predictive control, Monte Carlo Decision Trees (MCDTs), A* search, or various forms of look-ahead approximations and dynamic programming.

We argue that this predictive capability requires coupled digital twins of the instrument and the sample. Following the National Academies definition, a digital twin is a dynamically updated computational representation of a physical system that supports prediction and decision-making[15,19-22]. For scientific instrumentation, the instrument twin represents signal formation, feedback dynamics, operating limits, and measurement uncertainty. The sample twin, in turn, represents prior knowledge, experimental observations, and evolving beliefs about material state and behavior. Their coupling gives the planner a forward model of the experiment rather than a record of past measurements alone.

This dual view also natively defines calibration and discovery. Calibration reduces uncertainty in the instrument twin by using a known sample, while discovery reduces uncertainty in the sample twin by using a calibrated instrument. In both cases, measurements update the coupled state and improve prediction. The same framework therefore links instrument calibration, experiment design, adaptive sampling, and autonomous exploration.

Scanning probe microscopy (SPM) is a stringent test case for this idea because its observables are not produced by the sample or instrument alone. Instead, measured signals arise from sample properties, tip-sample interactions, cantilever dynamics, feedback control, and instrument electronics acting together[23-28]. Existing physics-based simulators capture important parts of cantilever and interaction physics[29,30]. Machine-learning models, by contrast, provide flexible empirical prediction[3-5,7,31-33]. Yet neither approach alone gives a calibrated, continuously updateable representation for predictive operation. Physics models are often difficult to parameterize for a specific instrument state, while purely learned models often lack interpretability and reliable extrapolation.

Here we realize the predictive layer of a coupled digital-twin framework for amplitude-modulation scanning probe microscopy. We combine a physics-informed representation of force–distance measurements, a deterministic scanner model describing cantilever and feedback dynamics, and sparse learned residual corrections that capture instrument-specific deviations from the physical model. The resulting framework provides a calibrated forward model capable of predicting scan quality, force-based safety metrics, and measurement outcomes across operating conditions. While the realization of fully autonomous experimental planning remains a future objective, the present work establishes the predictive foundation upon which such decision-making systems can be built.

## II. Coupled digital twins for predictive microscopy

The central concept of this work is the interaction between an experimental planner and a pair of coupled digital twins. The planner may be either a human operator or an AI agent, and its role is to translate experimental intent into candidate sequences of microscope operations, including choices of imaging modes, operating parameters, measurement locations, and acquisition workflows. The role of the coupled sample and instrument twins is fundamentally different. Rather than selecting actions, they predict the likely outcomes of candidate actions together with associated uncertainties and risks. These predictions can be used directly to construct acquisition functions for optimization, estimate measurement safety, evaluate expected information gain, or, more generally, provide the roll-out functions required for long-horizon experimental planning. In this sense, the planner proposes actions while the coupled twins evaluate them. The present work focuses on realizing the predictive layer represented by the sample and instrument twins, establishing the foundation required for future autonomous planning and decision-making systems.

Figure 1 makes this separation of sample and instrument twins explicit. A digital twin is useful for autonomous microscopy only when it can evaluate candidate actions before they are taken. To make that evaluation possible, the instrument twin estimates how the microscope will respond. At the same time, the sample twin estimates how the material state constrains that response and how it may evolve. The planner then combines predicted outcomes of possible actions to choose actions under uncertainty. Based on actual experiment, the twins are updated with the belief balance set by experiment goal and model details. This diagram sets the logic for the rest of the manuscript: define the two twins, couple them, test their joint predictions, and examine how the physical model must be corrected.

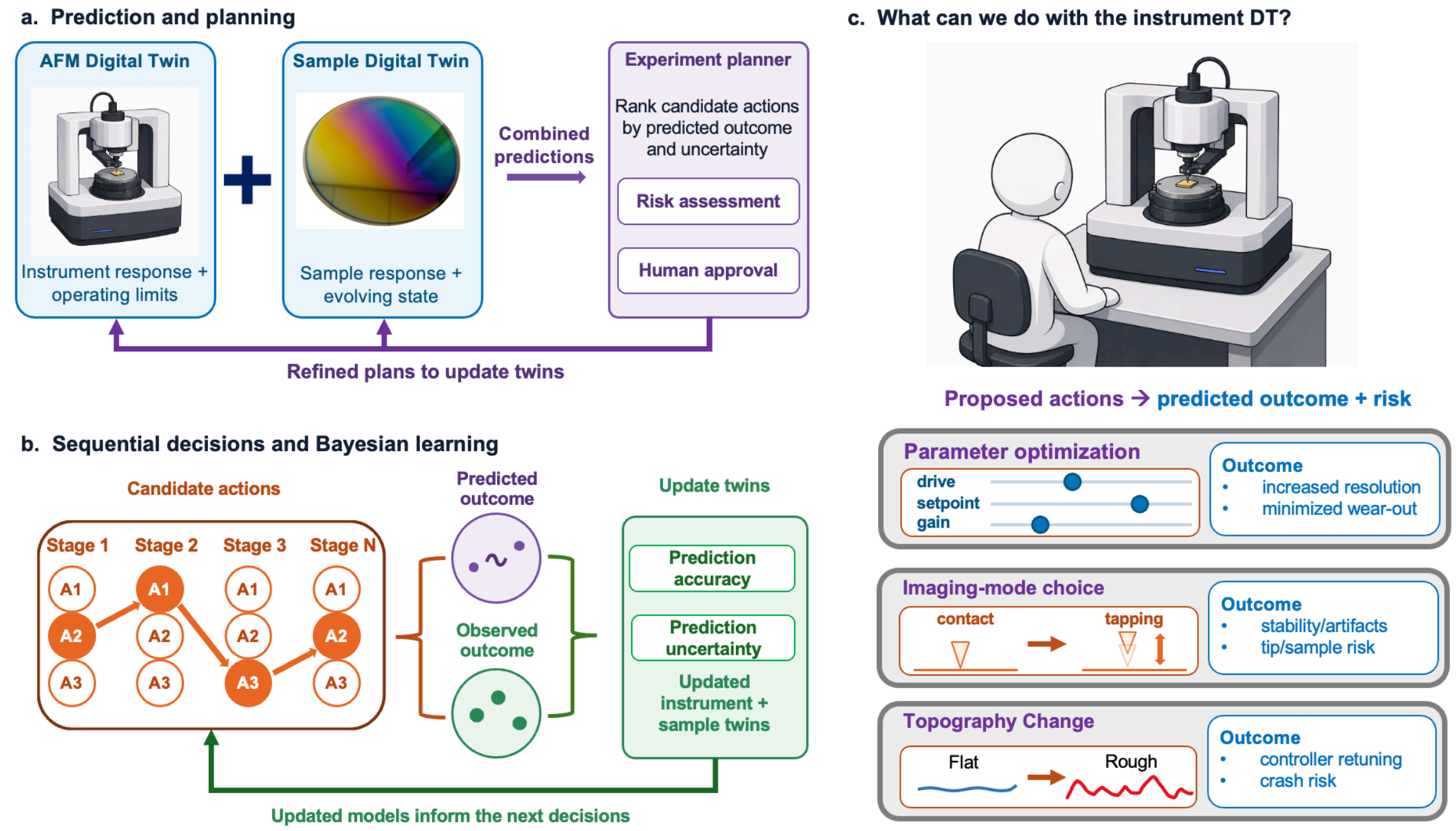

**Fig. 1 Coupled digital twins as the predictive layer for autonomous microscopy.** a, The experimental planner uses coupled sample and instrument digital twins to evaluate candidate actions before execution. b, The planner proposes candidate operations across an experimental workflow, while the twins predict expected outcomes, uncertainty, and risk. The observed outcome is then compared with the prediction to update the twins. c, The instrument twin estimates the consequence of candidate operating choices, supports risk-aware parameter selection, and helps interpret the resulting measurements.

### a. Sample twin

A central role of the sample digital twin is to represent, in machine-actionable form, what is known about the sample before and during experimentation. While scientific measurements are often framed as exploration of an unknown system, experiments rarely if ever begin from a state of complete ignorance. Information about material composition, processing history, intended functionality, prior characterization, and related material systems provides constraints on plausible behaviors. The purpose of the sample twin is to encode this information as a structured representation that can be updated as new measurements are acquired.

In the most general sense, the sample twin can exist at multiple levels of complexity. For exploratory measurements it may consist of broad priors over expected material responses. For mature systems it may incorporate detailed physical models and previously acquired experimental observations. Recent advances in large language models and reasoning systems further expand the range of information that can be incorporated into such priors, enabling the conversion of heterogeneous textual, experimental, and theoretical knowledge into machine-actionable constraints relevant to specific experiment. Here, we highlight that one of the key tasks is conversion of the total body of prior knowledge to the state variables relevant to the specific experiment. For example, for SPM in general expected roughness and mechanical properties are highly relevant, whereas chemical composition and color are not.

This general idea must be further refined and reduced to quantities that can be measured and updated during an experiment. For amplitude-modulation SPM, we realize the sample twin as a mechanical descriptor-based representation initialized from force-distance (FD) measurements and the instrument twin as a calibrated scanner model. Figure 2 shows how the two are coupled. FD curves first provide physical descriptors. The encoder then recovers those descriptors from measurements, the scanner propagates them through the feedback loop, and a learned residual layer corrects the remaining microscope-specific error. This architecture matters because every learned block is anchored to a physical state or a physical prediction, rather than to an uninterpretable latent vector alone.

### b. Instrument twin

The instrument twin is responsible for predicting how the microscope transforms sample properties into measured observables. In scanning probe microscopy, this transformation depends on the coupled dynamics of the probe, cantilever, excitation waveforms, detection chain, feedback controller, and signal-processing electronics. The purpose of the instrument twin is not merely to reproduce measurements, but to provide a predictive representation that can estimate the consequences of candidate operating conditions before measurements are performed.

Different components of the instrument possess different levels of observability and stability. Probe geometry and contamination state evolve during operation and must be inferred indirectly from measurements, with manufacturer specs providing required priors. Cantilever properties are generally more stable, though they may vary with environmental conditions and probe history. Feedback gains, operating modes, and excitation parameters are directly accessible control variables. Internal electronics and controller implementations are often only partially observable and must be represented through calibration and system identification.

These differences in observability determine the practical design of the instrument twin. We therefore use a deterministic scanner model that combines cantilever dynamics with a calibrated feedback controller. In the practical implementation here, the model maps FD descriptors from the sample twin to experimentally observable scan outputs. Figure 2 illustrates this concept and shows why the order of the steps matters. The workflow starts from FD curves and scan traces, converts them into physical descriptors, and passes those descriptors through a deterministic scanner. Only after this physical prediction is made does the framework learn the residual correction needed for real instrument behavior. The physical model therefore defines the coordinates and the expected response, while learning is reserved for the part of the response that the physical model misses. This ordering keeps the model tied to measurable physics while still allowing adaptation to microscope-specific effects.

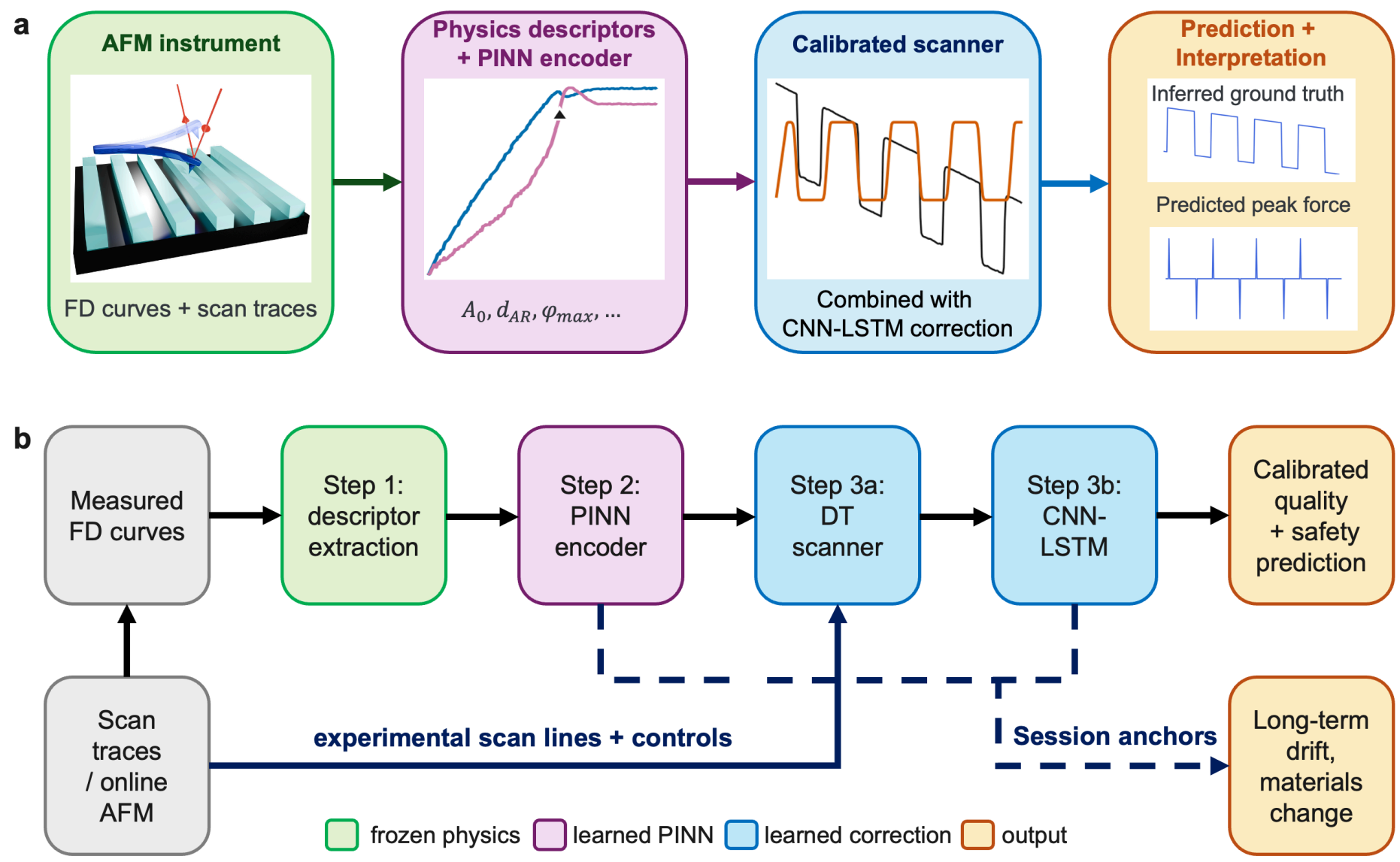

**Fig. 2 Descriptor-aligned digital-twin framework for amplitude-modulation SPM.** a, The microscope supplies force-distance (FD) curves and scan traces. A physics-informed encoder converts FD curves into physical anchors that reconstruct the height, amplitude, and phase response. These anchors parameterize a deterministic scanner that predicts scan quality and force-based safety. A CNN-LSTM residual model then corrects the remaining instrument-specific error, giving predicted scan quality, inferred topography, and peak-force-derived safety. b, The five-step workflow links descriptor extraction, PINN-based sample-state inference, deterministic scanning, short-term residual correction, and long-term drift monitoring.

## III. Descriptor-based sample twin

With the architecture defined, the next task is to specify the sample state that will be passed to the instrument twin. In principle, this state could contain detailed physical models, prior measurements, and evolving beliefs about material behavior. For online operation, however, such a representation must stay compact enough for rapid calibration and update. The key challenge is therefore to preserve the information that controls microscope behavior while reducing the dimensionality enough to learn from limited data.

FD measurements provide that starting point for amplitude-modulation SPM. FD curves probe the interaction between the oscillating cantilever and the sample, so they encode the material response that governs imaging behavior. For a new sample, an FD curve can be estimated from partially known materials information and then refined by measurement. Yet the full curve is too high dimensional to use directly in a fast scanner twin. We therefore seek a compact descriptor vocabulary that retains the parts of the interaction most relevant to operation.

### a. Physics-informed force-distance representation

The descriptor vocabulary is built from the physical model rather than chosen only from data. Specifically, the sample twin is based on a model of cantilever dynamics under a distance-dependent force field. The driven cantilever is represented as a damped nonlinear oscillator, whose steady-state amplitude and phase depend on tip-sample separation and interaction parameters. By integrating the equation of motion over separations and operating conditions, we build a library of FD responses that links measured observables to physical interaction descriptors.

This representation has two roles in the coupled twin. First, it gives an interpretable description of the interaction process. Second, it defines the coordinate system used to transfer observations across measurements. The goal is therefore not to reproduce every detail of each measured FD curve. Instead, the goal is to identify the quantities that control later microscope behavior and can be passed forward to the scanner model.

### b. Descriptor vocabulary as sample state

Figure 3 defines the descriptor vocabulary used as the sample state. The descriptors include the free oscillation amplitude, characteristic amplitude crossings, phase extrema, and the attractive-repulsive transition point. Together, these quantities summarize the interaction between probe and sample in a compact state. The same figure then connects the descriptors to scan quality and safety, which are the operational quantities the twin must predict. This link is important because the descriptors are not chosen only to fit curves well. They are chosen because they sit between material response and microscope decisions.

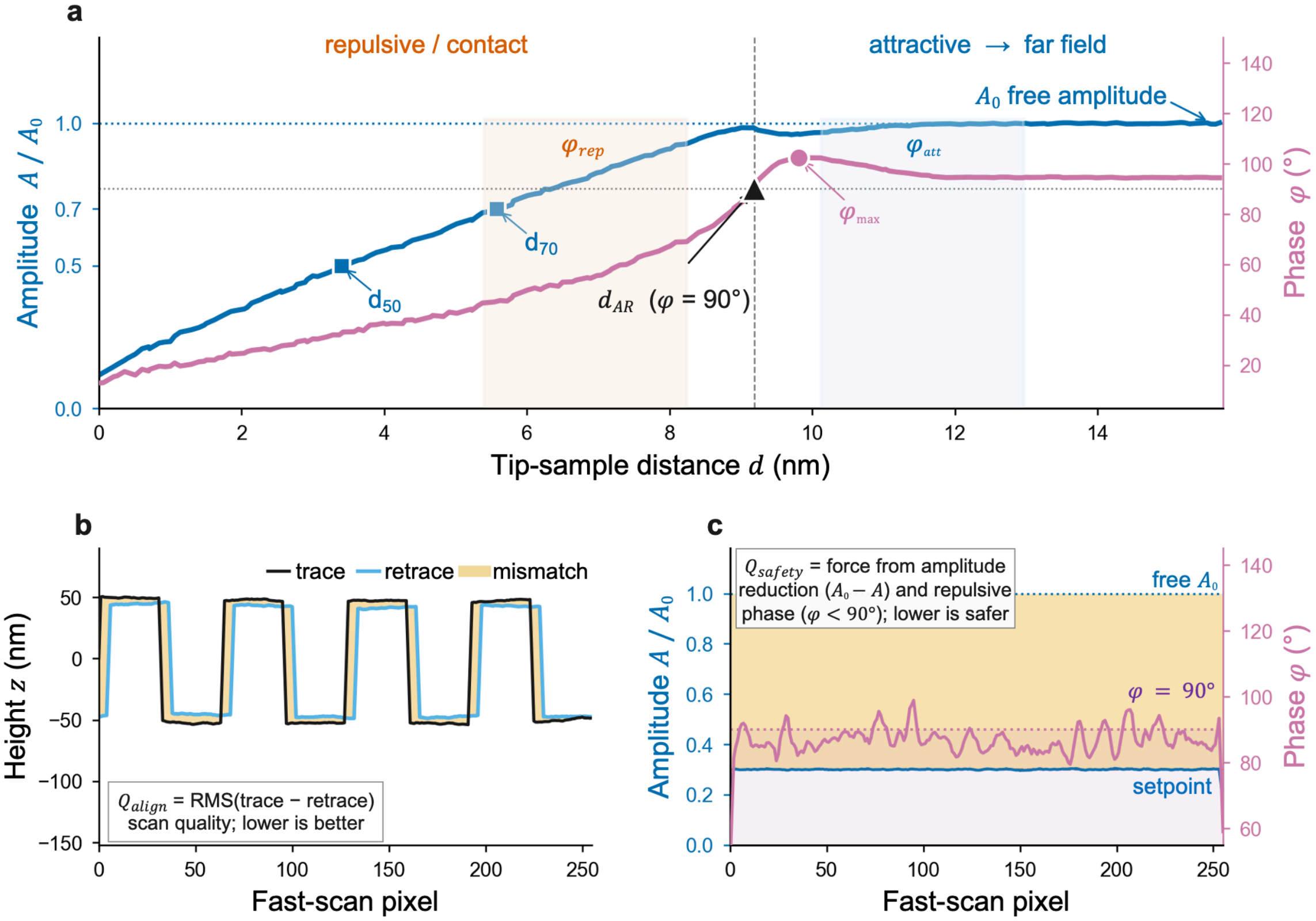

**Fig. 3 Physical descriptors and operational quality metrics.** a, FD descriptors extracted from a measured Tap-300 curve, including free amplitude, characteristic amplitude crossings, phase extrema, and the attractive-repulsive transition marked by the 90-degree phase crossing. b, The scan-quality metric $Q_{quality}$ is defined from trace-retrace mismatch in the height channel. c, The safety metric $Q_{safety}$ is derived from amplitude reduction and phase branch information. Amplitude reduction reports increasing force, while phase below 90-degree marks repulsive operation.

The descriptor set is intentionally restrictive because a useful twin must generalize beyond one measured curve. It acts as a bottleneck that keeps only information relevant to scanner operation as state variables. As a result, FD curves with the same descriptor values are equivalent from the scanner's perspective. Differences outside the descriptor set are discarded by design. This choice sacrifices completeness for transferability and interpretability. Because each descriptor has direct physical meaning, changes in the twin can be checked against measurable changes in the tip-sample interaction.

### c. Learning the sample state from measurements

After defining the descriptor state, the remaining sample-side problem is to infer it from experimental data. We infer these descriptor values with a physics-informed neural network (PINN)[34,35]. The network combines measured amplitude and phase curves with prior knowledge from the FD model. It then receives FD measurements and experimental metadata and predicts the descriptor set required by the instrument twin.

Figure 4 provides the first performance test of this construction. If the encoder cannot recover the physical anchors, the downstream scanner twin cannot be trusted. The held-out parity and error panels therefore test the link between measured FD curves and the state variables used for prediction, before any scanner prediction is evaluated.

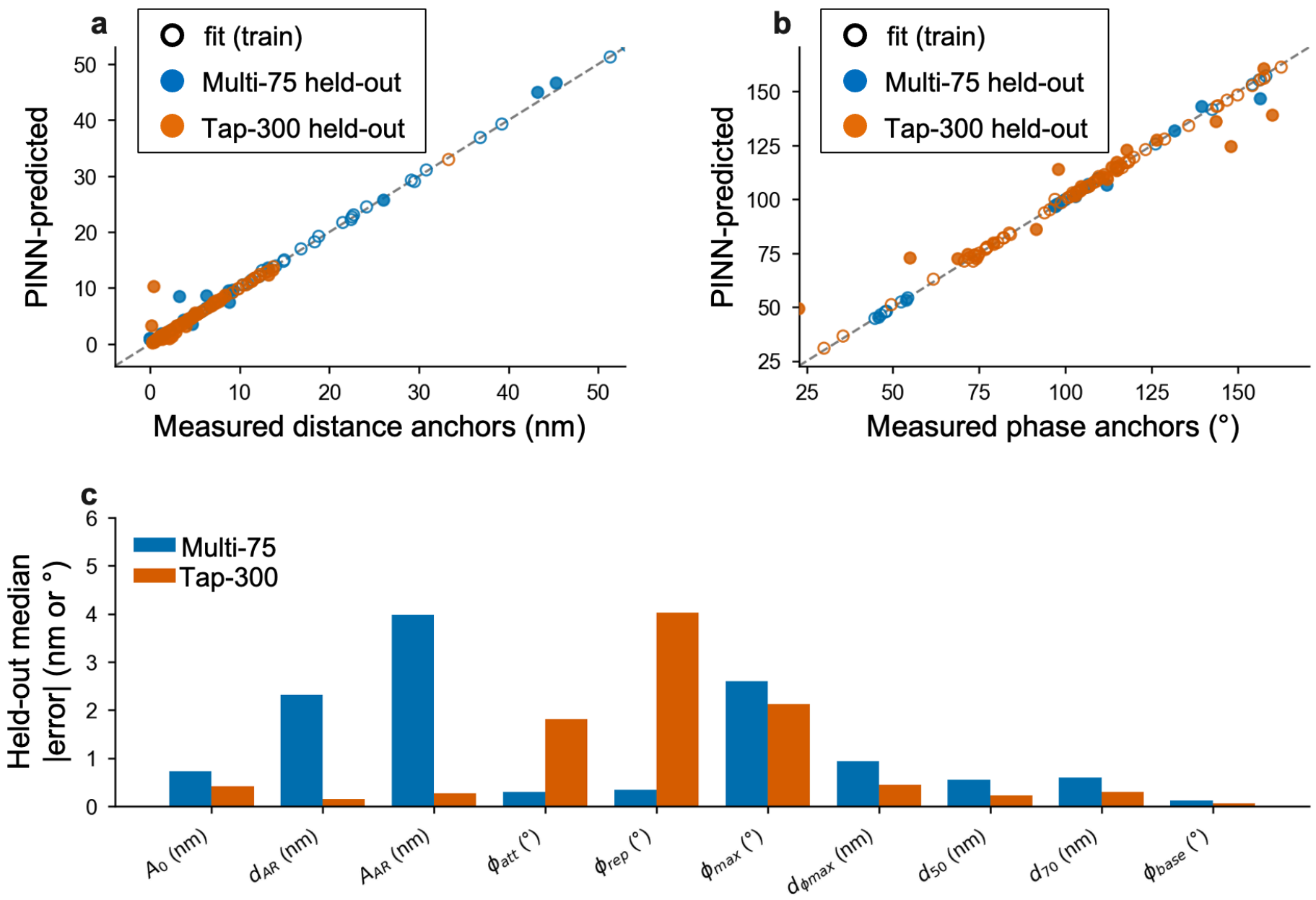

**Fig. 4 Physics-informed encoder recovers scanner-driving descriptors in physical units.** a, Predicted versus measured distance anchors for training curves and held-out curves. b, Predicted versus measured phase anchors. Points on the diagonal indicate exact recovery. c, Held-out median error for each anchor in physical units. The setpoint anchor is recovered to 0.4 nm, showing that the descriptor state can be inferred accurately enough to drive the instrument twin.

The results in Fig. 4 show that the proposed sample state can be learned with enough accuracy to drive the scanner twin. On held-out FD curves, the predicted descriptors closely follow experimentally determined values. In particular, the attractive-repulsive transition point is recovered to about 0.4 nm, and the free oscillation amplitude is recovered with sub-nanometer accuracy. This performance means that the descriptor vocabulary is not only interpretable, but also experimentally accessible. It can therefore serve as a practical interface between a real microscope and a predictive model.

The residual error is also informative because it shows where the current physical model reaches its limit. The largest errors occur in phase-related descriptors and contact-side interactions, where the single-mode cantilever model neglects higher flexural modes, torsion, and viscoelastic sample response. The supplementary analysis localizes this residual to the phase channel. Consistent with that diagnosis, rebuilding the phase from a one-field dissipation model lowers the held-out median phase error by 76% on the grating and 82% on AlScN. The implication is useful for model development: the amplitude channel can remain a reliable physical anchor, while phase requires an explicit dissipation or missing-physics term. The remaining discrepancy is therefore not just a prediction error. It marks the boundary between the current physical representation and the experimental system.

This stage produces the input required for the next stage: a calibrated sample twin. Its compact physical descriptors can now be passed to the instrument twin, which predicts microscope behavior under specific operating conditions. The descriptors therefore form the interface between material state and instrument response.

## IV. Instrument twin

The next step is to propagate the learned sample state through the microscope. The instrument twin performs this step by predicting how a sample state is transformed into experimentally observable signals. The sample twin encodes the material through descriptors. The instrument twin then converts those descriptors into measurements through cantilever dynamics, feedback control, and signal processing. In this way, it links material state to the observables available to the microscope.

This transformation is dynamic, which is why a simple map from sample descriptor to image contrast is not sufficient. Height, amplitude, and phase channels are not direct maps of sample properties. Rather, they are outputs of the cantilever, tip-sample interaction, feedback controller, and electronics acting together. Predicting microscope behavior therefore requires a model of the instrument dynamics.

### a. Deterministic z-scanner model

The instrument twin is implemented as a deterministic scanner model that combines the descriptor-informed force–distance response with a proportional–integral feedback loop. The scanner receives the descriptors extracted by the sample twin and reconstructs the effective amplitude–distance and phase–distance relationships associated with the current probe–sample system. These reconstructed responses are then coupled to a model of the feedback controller to simulate the evolution of the microscope signals during scanning.

This construction reflects a central design principle of the framework, in which the sample twin and instrument twin are explicitly coupled. The sample twin defines the interaction state, while the instrument twin defines the measurement process. Together they generate the predicted observables that would result from a particular experimental condition. The deterministic scanner reproduces the principal features of microscope operation across a broad range of operating conditions. For typical imaging regimes, predicted traces agree closely with experimental observations, demonstrating that much of the instrument behavior can be captured using a relatively compact physical model parameterized by the descriptor representation.

### b. Limits of the deterministic twin

A useful digital twin must do more than predict successful operation or risk. It must also identify where its assumptions are likely to fail. Figure 5 exposes this point by comparing deterministic scanner predictions with measurements over a wide range of operating conditions. The comparison shows strong agreement in many regimes. However, a distinct failure regime appears near light tapping, where the experimental signal becomes highly variable while the deterministic scanner remains stable. This mismatch points to missing physics rather than a simple fitting error.

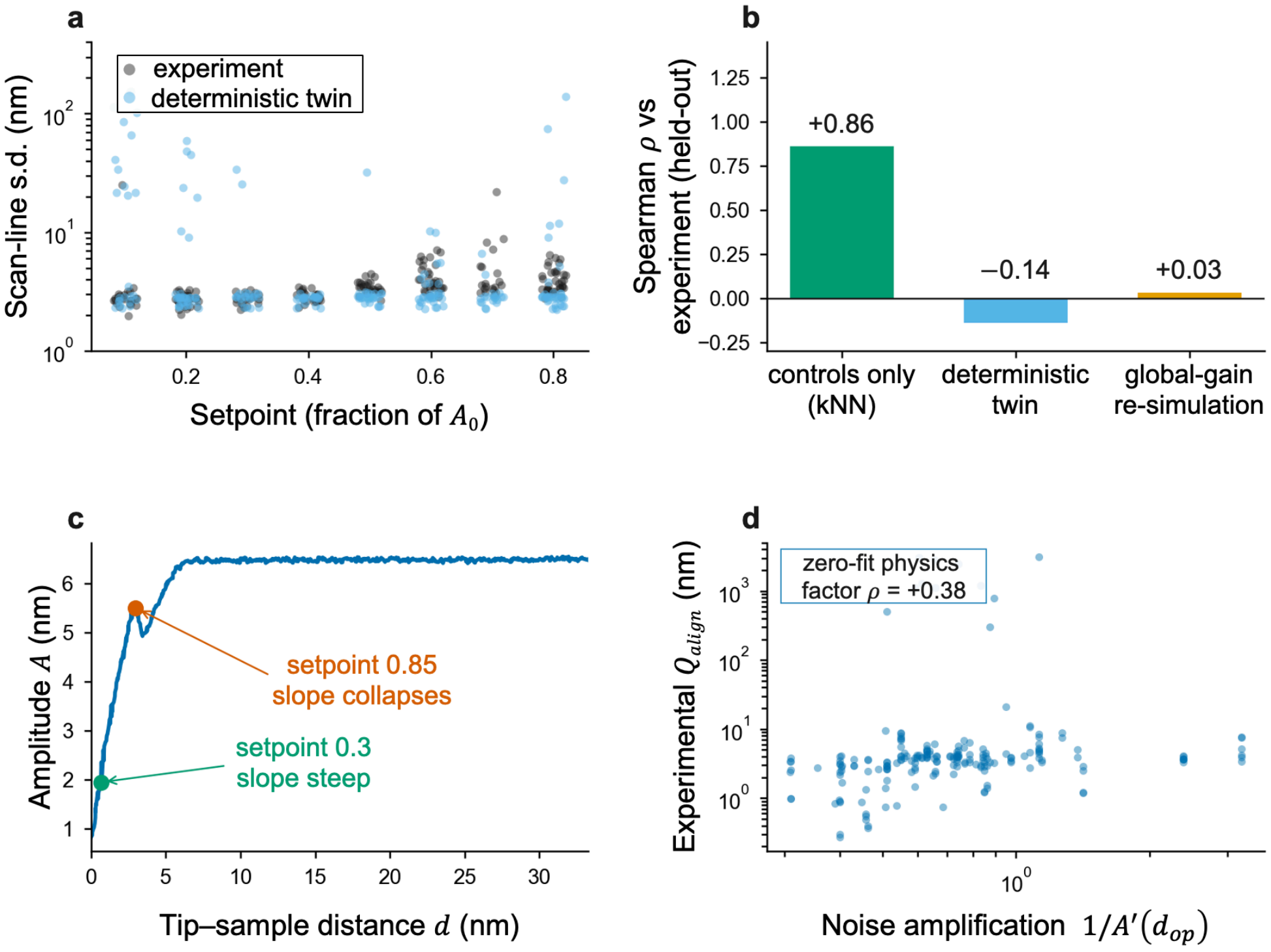

**Fig. 5 Deterministic scanner fidelity gap reveals an operating-point failure mode.** a, Held-out scan-line standard deviation versus setpoint shows that the experiment becomes unstable near light tapping, while the deterministic twin remains stable. b, Rank correlation with experiment for a control-only predictor, the deterministic twin, and a global-gain re-simulation. c, A measured FD curve shows that the local amplitude-distance slope collapses near the free amplitude. d, The zero-fit amplification factor $1/A'(d)$ at the operating point $d_{op}$ tracks held-out experimental quality, identifying noise amplification as the dominant missing-physics coordinate.

The FD response explains why this regime is difficult. Near the free oscillation amplitude, the local amplitude-distance slope becomes small. Because the feedback loop acts on measured amplitude, this small slope amplifies measurement noise and controller perturbations. The result is an operating-point instability: the same sample can look reliable or unstable depending on where the controller sits on the FD curve. This finding changes how the figure should be read. The deterministic twin does not merely fail near light tapping. Instead, it identifies the physical coordinate in which the failure becomes predictable.

Importantly, this effect can be identified directly from the force–distance library without introducing additional fitting parameters. The resulting operating-point amplification factor correlates strongly with experimentally observed degradation in scan quality and therefore provides a physically meaningful explanation for the dominant failure mode of the deterministic scanner. This result shows why the instrument twin is useful even when it is imperfect. Beyond producing predictions, it explains prediction failures and separates behavior captured by known physics from behavior that needs additional modeling.

### c. Instrument state, uncertainty, and missing physics

The observed failure mode also highlights the limitations of the present deterministic representation. The current scanner assumes idealized controller behavior operating on noise-free lock-in signals. Real instruments contain additional layers of complexity, including finite lock-in bandwidth, digital filtering, detector noise, and piezoelectric nonlinearities. Many of these effects are proprietary and are therefore only indirectly observable.

Consequently, certain forms of behavior cannot be reproduced by the current physical model alone. Rather than attempting to absorb these discrepancies into increasingly complex hand-built models, we treat them as structured residuals that can be learned from experimental data. This observation motivates the final component of the framework: a learned correction layer that operates on top of the calibrated sample and instrument twins.

The deterministic scanner therefore serves two complementary purposes. First, it provides a physically interpretable prediction of microscope behavior. Second, it establishes a baseline against which missing physics can be identified, quantified, and ultimately incorporated through data-driven correction. Overall, the deterministic scanner forms the core of the instrument twin, while the learned correction provides a mechanism for adapting the twin to the realities of a specific microscope operating under real experimental conditions.

## V. Predictive operation

With both twins constructed in terms of state variables and interaction mechansism and initialized, the framework can now be used for prediction. The combined model can be queried before measurement to estimate expected outcomes, uncertainty, and risk. Here, we focus on scan quality and measurement safety. These two quantities capture the central practical trade-off in microscope operation. Scan quality measures how faithfully structural information is acquired. Safety, in contrast, measures how strongly the probe-sample system is perturbed. Together, they describe the trade-off that an operator or autonomous planner must manage when choosing setpoint, gain, drive, and scan speed.

### a. Predicting measurement quality

The first operational task is to predict scan quality (Fig. 6). We define quality with descriptors derived from trace-retrace agreement and related height-signal characteristics. These metrics depend strongly on feedback-controller behavior. They therefore directly test whether the instrument twin converts the sample state into useful imaging predictions.

The deterministic scanner captures broad quality trends across operating conditions, but the failure analysis above explains why it cannot be sufficient everywhere. Discrepancies appear in regimes where noise amplification and unmodeled electronics dominate. Those discrepancies are precisely the part of the prediction assigned to the learned correction layer.

Figure 6 therefore moves from model construction to operational prediction. At this stage, the question is no longer whether the twin can reproduce internal descriptors, but whether it can predict quantities an operator needs before imaging: expected quality and force-related safety.

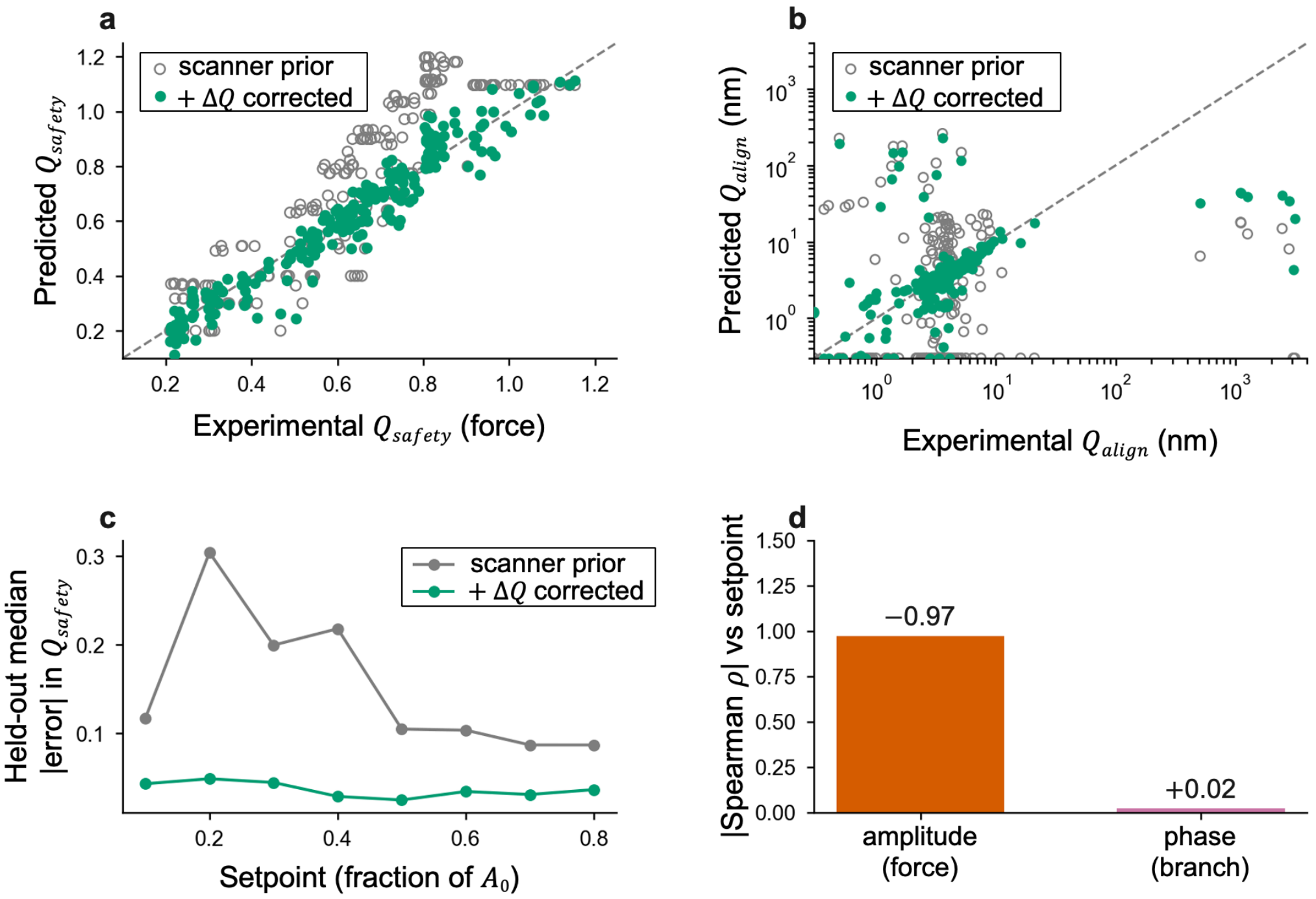

**Fig. 6 Learned residual correction enables predictive quality and safety estimates.** a, Tap-300 force-based $Q_{safety}$ parity for the deterministic prior and corrected prediction. b, Tap-300 $Q_{quality}$ parity on a log-log scale after correction. c, Held-out median $Q_{quality}$ error versus setpoint shows that the correction maintains low error across the tapping regime. d, Amplitude and phase provide complementary information: amplitude tracks force and setpoint, while phase identifies the interaction branch. Keeping these channels separate enables prediction of both scan quality and force-related safety before acquisition.

The correction model operates on outputs from the calibrated twins rather than on raw experimental measurements. It therefore learns residual behavior not explained by the physical model. Because the residual is learned after the physical prediction, the architecture preserves interpretability while improving accuracy. Across held-out conditions, the corrected model reduces typical scan-quality prediction error by more than an order of magnitude relative to the deterministic baseline. Figure 6 also explains why amplitude and phase should not be collapsed into a single empirical channel. Amplitude carries force information and follows setpoint, whereas phase identifies the interaction branch. Keeping these channels physically separated lets the model predict both quality and safety before the scan is run.

### b. Predicting measurement safety

The second operational task follows naturally from quality prediction. A high-quality image is not sufficient if the probe or sample is damaged during acquisition. Scan quality is derived mainly from height measurements, whereas safety is tied to tip-sample forces. We therefore evaluate safety using descriptors from amplitude and phase signals, not topography alone. The safety metric captures both interaction magnitude and interaction regime. Amplitude reduction reports increasing force, while phase distinguishes attractive and repulsive operation. Together, these channels provide a physical measure of measurement risk.

The deterministic scanner already predicts the safety metric well because the metric is closely linked to the FD response encoded in the sample twin. The learned correction further improves quantitative accuracy and yields strong agreement between predicted and observed safety values across the operating range. This behavior is consistent with the design of the framework: when a target quantity is controlled by physical descriptors, the physical anchor carries much of the prediction.

For autonomous operation, this result is important because it makes risk a predictive variable rather than a post-experiment diagnosis. Physically meaningful safety metrics can therefore be estimated from the coupled twins before measurements are performed.

### c. Physics, learning, and hybrid prediction

The quality and safety results raise a broader question: how should physics and machine learning share responsibility in a scientific digital twin? Figure 7 addresses this question by comparing pure physics, pure data-driven prediction, and a hybrid residual model. In this comparison, the sample and instrument twins provide a calibrated physical representation of microscope and sample state. The learned correction captures deviations caused by incomplete physics, unmodeled instrument behavior, and experimental variability. The balance between these components depends on the target quantity.

Figure 7 explains why the framework should be hybrid rather than purely physical or purely data-driven. The physical model supplies the anchor, while the learned model supplies the missing detail. Because the correction is formulated as a residual, the boundary between the two remains visible. This boundary is the practical reason to use physical models as anchor points.

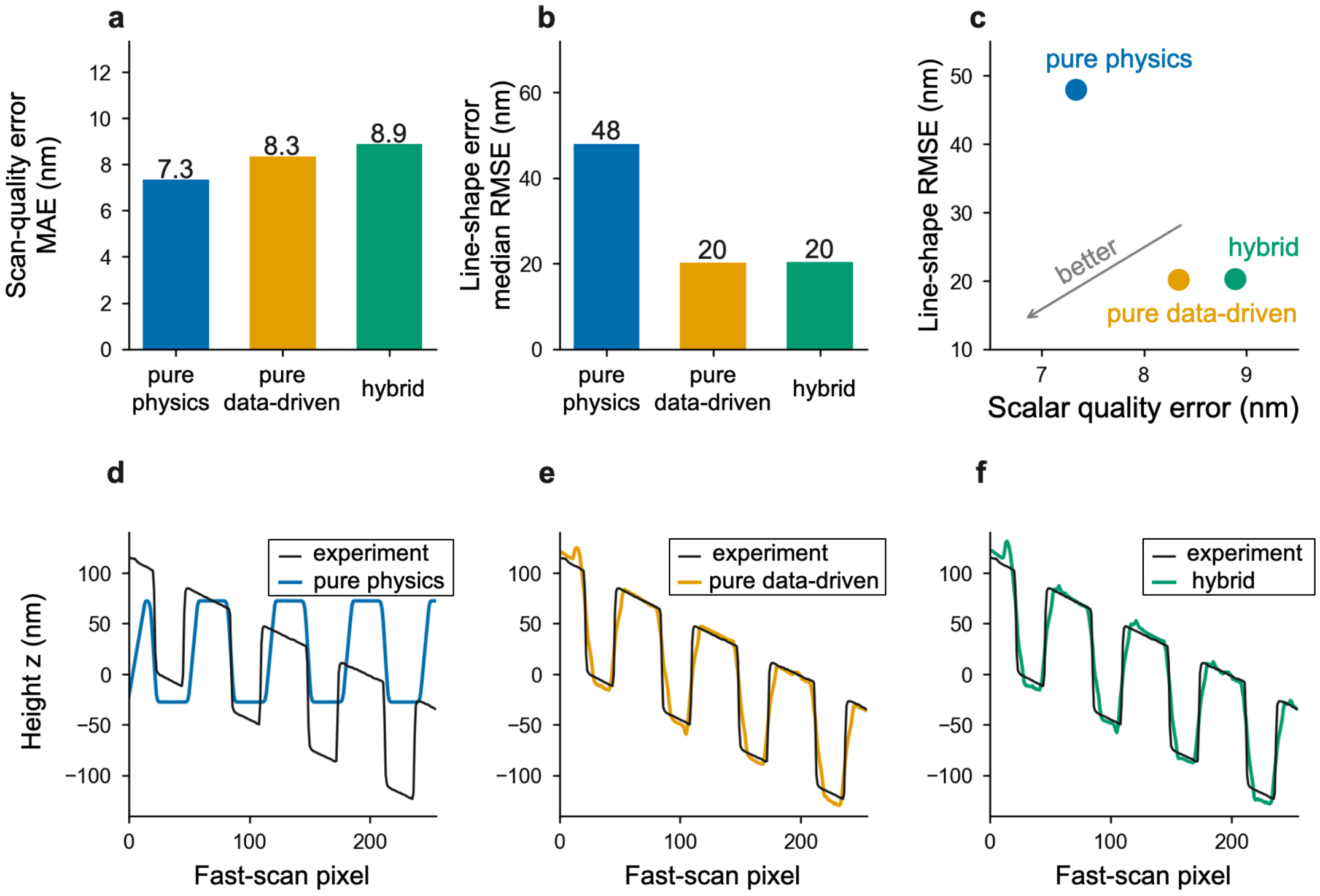

**Fig. 7 Physical anchors and learned residuals play complementary roles.** a, Held-out scalar scan-quality error for pure physics, pure data-driven prediction, and the hybrid model. b, Held-out line-shape error for the same models. c, The Pareto trade-off shows that no single model is best for both scalar quality and detailed line shape. d-f, A held-out grating line compares experiment with pure physics, pure data-driven prediction, and the hybrid model. Physics captures the dominant period but misses detailed shape, while learned models recover the line profile. The hybrid strategy preserves physical interpretability while adding data-driven flexibility.

To make this division explicit, we compare three strategies. The first uses only the deterministic sample and instrument twins and represents a physics-based approach. The second ignores the physical model and learns directly from data, while the third adds a learned residual correction to the calibrated physical twin. Together, these strategies represent mechanistic prediction, empirical prediction, and hybrid prediction.

The comparison shows a clear division of labor. For scalar quantities such as scan quality and safety, the physics-based twin performs well because the quantities are strongly constrained by instrument dynamics. The calibrated scanner therefore captures the dominant relationships between operating conditions and observables with relatively little data. Detailed scan-line reconstruction benefits more from learning-based models, which capture effects not included in the physical model. This result gives the central reason to use physical models as anchor points. They provide the correct coordinates, constraints, and failure diagnostics, while learning supplies flexibility where the physics is incomplete.

The hybrid model combines these strengths. The physical twin sets the operating regime and gives a meaningful baseline. The learned correction then captures systematic deviations from that baseline. As a result, the hybrid model achieves high predictive accuracy while preserving interpretability. It can explain why a prediction was made, identify the physical origin of failures, and remain useful where a purely data-driven model would need more training data. In practice, this combination means the coupled twin can support parameter selection, risk-aware imaging, and simulated roll-outs for future autonomous planners.

More broadly, the question is not whether physics or machine learning is better, but how the two should be assigned. Physical models are best suited for constraints, conservation laws, controller dynamics, and signal formation. Learning is better suited for residual effects that are difficult to model directly or only partly observable. In a digital twin, the learned component should therefore represent the uncertainty and incompleteness that remain after the physics has been used. This assignment is what makes the model both predictive and inspectable.

## VI. Conclusions and outlook

The transition from closed-loop optimization toward open experimental decision-making requires the ability to predict the consequences of candidate actions before they are executed. We have argued that this capability requires coupled digital twins of both the instrument and the sample. The instrument twin captures signal formation, feedback dynamics, and operational constraints, while the sample twin captures prior knowledge and evolving beliefs about material state. Calibration and discovery emerge as complementary limits of the same framework:

calibration reduces uncertainty in the instrument twin using a known sample, whereas discovery reduces uncertainty in the sample twin using a known instrument.

Here we have realized the predictive layer of such a framework for amplitude-modulation scanning probe microscopy. A physics-informed encoder recovers the descriptors linking force–distance measurements to instrument operation, a deterministic scanner model provides a calibrated representation of cantilever and feedback dynamics, and sparse learned residuals capture instrument-specific behavior not represented by the physical model. Together these elements enable prediction of scan quality, force-based safety metrics, and measurement outcomes across operating conditions. The resulting framework also identifies the dominant source of model mismatch, here noise amplification near the feedback operating point, and demonstrates a clear division of labor between physics and learning: calibrated physics provides the most reliable prediction of scalar performance metrics, learned models provide the most accurate reconstruction of detailed measurement outputs, and hybrid approaches combine the strengths of both.

More broadly, this work demonstrates that microscope operation can be formulated as the predictive use of a calibrated forward model. Parameters can be evaluated before acquisition, measurement outcomes can be compared against expectations, and residuals can be interpreted in terms of underlying physical mechanisms. While the present work focuses on the predictive layer itself, it establishes the foundation required for future autonomous experimental decision-making. Several immediate directions follow. On the instrument side, incorporation of stochastic scanner dynamics, realistic lock-in transfer functions, dissipation channels, and long-term drift models will extend predictive accuracy beyond the deterministic operating regime. On the sample side, future twins should evolve from descriptor-based representations toward continuously updated belief states capable of integrating information from multiple measurements and complementary characterization modalities.

The broader opportunity lies in moving from prediction to decision-making. Once calibrated, digital twins can evaluate candidate actions before execution, estimate force-based risk metrics, predict expected rewards, and quantify uncertainty. In this framework, comparison between predicted and observed outcomes naturally defines Bayesian surprise, providing a mechanism for identifying where uncertainty resides and which measurements are expected to be most informative. The same predictive capability enables simulation of candidate action sequences, providing the roll-out functions required by planning algorithms that operate over extended experimental horizons.

Realizing this vision will require both improved models and broader access to instrument knowledge. Many of the most consequential components of microscope behavior including controller firmware, lock-in transfer functions, detector characteristics, calibration tables, and piezo dynamics remain proprietary or inaccessible to end users. Instrument vendors therefore have an opportunity to expose calibrated controller and electronics modules that can be integrated into digital-twin frameworks while preserving intellectual property. In parallel, community-scale archives of force–distance measurements, scan histories, and instrument states could support foundation models spanning probes, samples, operating modes, and failure regimes. Together, these developments would transform digital twins from calibration tools into predictive engines for autonomous experimentation, enabling future systems to move beyond parameter optimization toward open experimental decision-making.

## Acknowledgements

YL, BS and SVK acknowledge support from the National Science Foundation Ceramics Program NSF 2523284. IM, and JPM acknowledge support from the Department of Energy Frontier Research Center (DoE ERFC) 3DFeM$^2$ program DE-SC0021118.

## Methods

The AlScN thin films are grown with a custom-built reactive RF magnetron sputtering chamber on 4 in. (100) n-type (As doped, 0.001 – 0.005 Ω*cm) Si wafers. Metal Al (99.9995 % Kurt J. Lesker) and Sc (99.9 MSE Supplies) 2 in. sputter targets are used with a Kurt J. Lesker RF 301 RF power supply and an ENI Odyssey DC power supply, respectively. The Si substrate is brought to 600ºC for 30 minutes then 350 ºC for 30 minutes, followed by a 50 W RF surface nitriding where 20 scccm Ar and 20 sccm $N_2$ are flowed into the chamber at 20 mTorr for 30 minutes. Then a 10 mTorr Ar-only presputter cleans the Al target at 200 W RF and the Sc target at 500 W constant DC. The deposition pressure is set to 1.9 mTorr and the Ar/$N_2$ gas ratio is 24/16 sccm. Stage bias is set to 2.5 W during the deposition, the Al power is increased to 300 W, and the substrate shutter is opened for 10 minutes to deposit on the wafer. Stage rotation is turned off during the deposition to achieve a composition gradient across the 4 in wafer.

**Datasets and held-out splits**

Two systems were used. Multi-75 was measured on a calibration grating with 15 FD curves and 60 scan conditions over drive, setpoint, and integral gain. These curves are far-field, so contact descriptors are undefined for all 15 curves. Tap-300 was measured on AlScN with 30 curves and 900 conditions over drive, setpoint, integral gain, and scan speed[6]. In contrast to Multi-75, these curves reach the attractive-repulsive transition on 29 of 30 curves. Curves were split 75/25 per system for Step 2. Scan conditions used the archive's predefined held-out set for Step 3. Seeds were fixed in both training cells, so before-and-after comparisons are not confounded by initialization.

**Scan-quality descriptors**

$Q_{align}$ is the root-mean-square trace-retrace difference of the height lines, calculated on centered interior pixels. $Q_{safety}$ is a force index read from amplitude and phase lines. At each pixel, the index is $(A_0 - A)/A_0$, the amplitude reduction below the free amplitude, plus a half-weighted repulsive term max(0, (90 degrees - $\varphi$)/90 degrees). The descriptor is the 90$^{th}$ percentile of this index along the line, averaged over trace and retrace. $Q_{align}$ uses the height channel because alignment is topographic. $Q_{safety}$ instead uses amplitude and phase because force information is encoded in those channels. The supporting height descriptors $Q_{stab}$, $Q_{grad}$, and $Q_{range}$ are unchanged. The same extractors run on experiment and on the twin.

**Step 2: physics-informed encoder**

A one-dimensional convolutional network reads the resampled amplitude and phase curve as two channels. It then concatenates the drive scalar and one-hot probe and sample tokens, and feeds a two-layer MLP into three heads: 18 anchors, five scanner parameters, and a per-(probe, sample) auxiliary table. The convolutions are 2 to 16 (kernel 7) and 16 to 32 (kernel 5), followed by adaptive average pooling to length 8. The loss combines a NaN-masked, tier-weighted descriptor term on training curves, an anchor-parameter consistency term, an ODE-residual proxy, and an L1 sparsity term that starts at epoch 200. Accepted defaults were weight decay of 1e-4 on both systems and ReLoBRaLo balancing on Multi-75 only. Random Fourier features and an uncertainty head were rejected on held-out evidence and remain optional.

**Step 3a: scanner wiring and the operating-point feature**

The encoder predictions for $A_0$ and $d_{AR}$ rescale and shift the deterministic scanner trace for each condition, with interpolation over drive on finite pairs. The five quality descriptors are then re-extracted. Two acceptance checks separate wiring integrity from fidelity. The wired scanner must preserve the reference scanner's ranking, and it must not lose agreement with experiment relative to the reference within the rank-correlation noise. The amplification feature is $1/A'(d)$ at the operating point $d_{op}$, where the amplitude crosses the setpoint multiplied by $A_0$. Conditions with no crossing are flagged and median-filled.

**Step 3b: learned reward correction**

$Q_{align}$ uses a per-line convolutional stem (2 to 16 to 32 to 64 to 32, kernel 9, stride 2) over the height window. An LSTM then runs over 32 conditions, is conditioned on the scanner prior, and passes its output to an MLP that emits the correction. Training runs in robust z-space using the median and interquartile range of the training targets, while reported metrics are in raw units. A Huber loss is the Tap-300 default because of its heavy tail. A ridge recalibration regresses the scanner prior on quadratic control features and the operating-point feature, fitted on training windows. Force-based $Q_{safety}$ uses the same window architecture on the two-channel amplitude and phase line, represented by amplitude reduction and the repulsive-phase index. Its deterministic prior is the steady force from the setpoint and the FD phase at the operating point. The correction is conditioned on the prior and the operating-point feature and is trained with Huber loss. Channel attention and synthetic augmentation were rejected on held-out evidence.

**Physics, data-driven, and hybrid predictors**

For this comparison, the same held-out conditions are predicted by three models: the deterministic scanner (pure physics), a model that maps controls and FD features to the centered line (pure data-driven), and a hybrid model that adds a learned residual to the physics baseline. Scan quality is the scalar trace-retrace metric. Line-shape error is the per-condition RMSE between aligned predicted and experimental lines. All three models use the same held-out set.